# THCluster:Herb Supplements Categorization for Precision Traditional Chinese Medicine

Chunyang Ruan*, Ye Wang†, Yanchun Zhang*†, Jiangang Ma†,
Huijuan Chen‡, Uwe Aickelin¶, Shanfeng Zhu*, Ting Zhang§
*School of Computer Scinece Fudan University, Shanghai, China
Email: cyruan16@fudan.edu.cn, zhusf@fudan.edu.cn
Telephone: (86-21) 51355059, Fax: (86-21) 51355558
†College of Engineering and Science, Victoria University, Melbourne, Australia
Email: yanchunzhang@fudan.edu.cn, ye.wang10@live.vu.edu.au, Jiangang.Ma@vu.edu.au
‡School of Basic Medical Science, Shanghai University of Traditional Chinese Medicine, ShangHai, China
Email: chenhuijuan70@sina.cn, zhangtingdoctor@sina.cn
¶School of Computer Science, The University of Nottingham, Ningbo, China
Email: uwe.aickelin@nottingham.edu.cn

*Abstract*—There has been a continuing demand for traditional and complementary medicine worldwide. A fundamental task in Traditional Chinese Medicine (TCM) is to optimize the prescription by mining knowledge from TCM to discover related categorization of herbs and effectively support diagnosis. In this paper, we propose a novel clustering model to solve this general problem of categorization. The model utilizes random walks, Bayesian rules and EM models to complete a clustering analysis effectively on a heterogeneous information network. We performed extensive experiments on the real-world datasets and compared our method with other algorithms and experts. Experimental results have demonstrated the effectiveness of the proposed model for discovering useful categorization of herbs and its potential clinical manifestations.

## I. Introduction

With the rapid development in science and technology, Traditional Chinese Medicine (TCM) plays an important role to improve people's health. TCM is an ancient medical practice science that emphasizes regulating the integrity of the human body and its interrelationship with natural environments [1]. Herbs are a fundamental part of TCM and their uses have to be conducted carefully to ensure the treatment effectiveness, safety assurance, and patient affordability. In general, western medicine make use of many 'one size fits all' non-prescription drugs for common diseases. In contrast, TCM has been using individual-based treatment as the core philosophy of clinical practice widely used in China for thousands of years.

In TCM clinical diagnosis, doctors observe their patients and record their symptoms. These symptoms are often captured and described in information entities such as 'throat-obstruction (喉痹)' and 'hemiplegia (半身不遂)'. According to these symptoms, doctors prescribe medicines, which have the functions to eliminate these symptoms. For example, the herb 'solidaginis herba (一枝黄花)' can treat 'throat-obstruction' because of its function of 'heat-clearing(清热)'. This 'suits the remedy of the case' manner in which TCM observes an individual's symptom patterns is reminiscent of precision medicine techniques. There is a general belief that TCM can complement modern precision medicine, which currently relies on molecular profiles [2].

TCM is an ancient, holistic treatment system established through empirical evaluation [3] and has accumulated a lot of knowledge for several thousands of years. Unfortunately, it is difficult for a TCM doctor especially young doctors to master the large amount of knowledge of herbs. Therefore, herb categorization can greatly increase our knowledge of medicine and assist doctors to prescribe a supplement in clinics. In addition, there are not many documents to record knowledge about disease diagnoses based on TCM-specific symptoms and herb prescriptions, which presents a significant challenge to effectively retain, share, and inherit knowledge of TCM [4].

To overcome the challenge, the technologies of machine learning and data mining can be used to discover knowledge from TCM. For example, Huang et al. [5] made inexact matchings of symptoms and herbs by using an embedded algorithm. Zhang et al. [6] propose a data mining method called the Symptom-Herb-Diagnosistopic (SHDT) model to find the relationship of symptom-herb-diagnosis from clinical data. Liu et al. [7] propose a new asymmetric probabilistic model for the joint analysis of symptoms, diseases, and herbs in patient records to discover and extract latent TCM knowledge. However, all these works require large-scale data such as (long texts) containing rich semantic information. Unfortunately, the majority of TCM corpora are lack of good semantics among herbs.

In this paper, we aims to find and discover relationship among herbs from TCM corpora without semantic information and structure. In particular, we study a novel clustering problem on a general Heterogeneous Information Network (HIN) and propose a novel algorithm THCluster to address the clustering problem found in the HIN. We design a novel star-schema network of TCM-HIN by utilizing entities and relationships from TCM corpora. Moreover, we use a path-based random walk method for the network to generate the reachable probability of entities, which can be effectively used to estimate the cluster membership probability and the importance of entities. THCluster can obtain the steady and clustering results through iteratively analysing network. We perform the experiments on real TCM datasets to validate the effectiveness of THCluster. The results show the accuracy and effectiveness of THCluste and also show excellent performance on expert opinions.

Our work has the following distinctive contributions:

- We present a clustering-based framework called THCLuster for herb categorization to discover the association of herbs, symptoms, diseases, formulas, functions. THCLuster can help doctors optimize TCM prescriptions. We use a path-based random walk method for the network to generate the reachable probability of entities, which can be effectively used to estimate the cluster membership probability and the importance of entities. In particular, THCluster can obtain the steady and consistent clustering results through iteratively analysing the network.
- We propose a novel cluster for TCM-HIN comprised of multiple types of enities. In each cluster, statistical information such as the member probability for each enity are derived to facilitate users to navigate in the cluster.
- We conduct an extensive experimental evaluation over a set of real datasets of TCM datasets. Our experimental results show that THCluster can give give quite reasonable clusteing results. The clustering accuracy is much higher than the baseline methods. THCluster also show excellent performance for optimizing prescription verified by TCM experts.

The rest of paper is organized as follows: Section II introduces the related literatures; Section III formally introduce several concepts and the clustering problem; Section IV systematically describes the THCLuster method; Section V presents the experimental design and discusses the results; Section VI concludes this study.

## II. RELATED WORK

In this section, we introduce related researches in HIN and the main work of cluster based algorithms in HIN.

As most real systems contain multi-typed interacting components, we can model the systems as heterogeneous information networks with different types of entities and links [8]. For example, in a medicine database like in TCM, herbs are connected together via symptoms, diseases, formulas and functions (an example is shown in Figure.1). Compared to widely-used graphs or networks, the HIN can effectively utilize more information such as topology, semantics in nodes and links, and thus it forms a new development in data mining. Recently, a growing number of researchers have noticed the importance of heterogeneous information network analysis (HINA) and apply HIN to solve more and more innovative data mining issues. According to the use of different methods, current research work can be classified into the following six categories: similarity search [9], [10]; **clustering** [11], [12]; classification [13]; Link Prediction [14]; ranking [15]; recommendation [16].

Conventional clustering is based on the features of entities, such as k-means and so on [17]. Conventional clustering algorithms are very mature for general problems, but their applicability is more limited in special data. For example, medical data usually is unstructured and considered to be one of the most difficult domains for data mining. HIN clustering needs to solve this difficult problem. Wang et al. use world knowledge for indirect supervision and study document clustering of multiple types of HIN with constraints [12]. Li et al. propose to generate clusters integrated with entity attribute values and their structural connectedness in the HIN and theoretical and experimental analysis showed that the quality of clustering is good [18]. Cruz et al. compose an attributed graph and use clustering to solve the community detection problem [19]. Zhang et al. propose a non-negative matrix tri-factorization multi-type co-clustering method called HMFClus to cluster all types of entities in HIN simultaneously [20]. Zhou designed a dynamic learning algorithm SI-Cluster to analyse social influence based graph clustering [21]. Luo et al. measure the similarity between two data entities and perform clustering with labelled information [22]. Boden et al. detect clusters by considering the connections in the network and the vertex attributes. Undoubtedly, these approaches improve the clustering performance effectively, but they are still confined to entities with rich attributes or labelled data. However, many real datasets are not perfect. Thus, we need more novel ideas for these particular datasets.

## III. PROBLEM DEFINITION AND FORMULATION

In this section, we present important concepts and problem formulation used in this paper.

The aim of this paper is to discover the category of a herb from the perspective of diagnosis. We employ $V^*$ and $E^*$ to denote the set of types of TCM entities and relations, respectively. In this paper, we have $V^* = \{H, S, D, F_m, F_c\}$ and $E^* = \{HS, HD, HF_m, HF_c, SF_m, DF_m, F_cF_m\}$, where $H$, $S$, $D$, $F_m$, $F_c$, represent these entities 'herbs', 'syndromes', 'diseases',' formula', and 'function' respectively, and $HS$, $HD$, $HF_m$, $HF_c$, $SF_m$, $DF_m$, $F_cF_m$ represent the relations herb-symptom, herb-disease, herb-formula, herb-function, formula-symptom, formula-disease and formula-function, respectively.

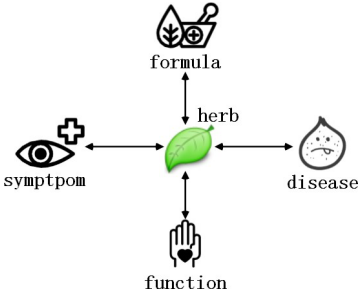

Fig. 1. An example of TCM-HIN schema network. The target type herb is marked with green color. Other types are attribute types, including symptom, disease, formula and function.

**TCM-HIN**: Let $I$ be a type of TCM types set $\{T\}$ and enity set can be defined as $V* = \{V^{(I)}\}$. The $V^{(I)} = \{V_q^{(I)}\}$ and $V_q^{(I)}$ represents an entity $q \in V^{(I)}$. The relations amongst entities can be denoted as $E* = \{E^{(I,J)}\}$ and $E^{(I,J)} = \{E_{q,p}^{(I,J)}\}$ where $E_{q,p}^{(I,J)}$ is the relation between a entity $q \in V^{(I)}$ and an entity $p \in V^{(J)}$. We define a TCM-HIN as a graph $G = \{V*, E*\}$, where $V* = \{V^{(I)}\}, E* = \{E^{(I,J)}\}$. Figure 1 shows a schema of TCM-HIN. TCM-HIN has five types of entities and becomes a star-schema network. The aim of this paper is to provide methods to discover the categories among herbs. So, we select herb types as the target type and connections to other types (symptom, disease, formula and function) as attributive types.

**TCM Heterogeneous Relation Matrix**: In order to utilize the relationships in TCM-HIN and calculate their values easily, we transform relations in TCM-HIN into adjacency matrices. The heterogeneous relation matrix can be denoted as $A^{(I,J)}$, where $A_{q,p}^{(I,J)}$ represents the relation between $V_q^{(I)}$ and $V_p^{(J)}$.

**TCM Heterogeneous Transition Matrix**: We define $H^{(I,J)}$ as the transition matrix, where $H^{(I,J)}$ can be derived from the relation matrix $A^{(I,J)}$ and $H^{(I,J)} = S^{(I,J)^{-1}} \times A^{(I,J)}$, where $S^{(I,J)^{-1}}$ is the diagonal matrix with the diagonal value equaling to the corresponding row sum of $A^{(I,J)}$. Taking Figure 1 as an example, $H^{(symptom, herb)}$ is the transition probability matrix of the 'symptom-herb' relation $A^{(symptom, herb)}$.

**TCM Member Matrix**: Clustering in TCM-HIN includes five types of entities and we give the member matrix for each type of entity $^{(I)}$. The member matrix can be denoted as a diagonal matrix $M^{(I|C_k)} \in [0, 1]^{|V^{(I)}| \times |V^{(I)}|}$ where the diagonal value represent the entity probability of $V_q^{(I)}$ belonging to cluster $C_k$. It is worth noting that the sum of all entity probabilities of $V_q^{(I)}$ in K cluster is 1 (i.e., $\sum_{k=1} M_q^{(I|C_k)} = 1$).

Now, we can formulate the key problem of clustering on the TCM-HIN as follows. Given the TCM-HIN $G = \{V*, E*\}$ and the cluster number K, the aim of this paper is to find a clusters set $\{C_k\}_{k=1}^K$, where $C_k$ is defined as $C_k = \{\{M^{(I|C_k)}\}_{I \in \{T\}}\}$. In this way, an entity $q$ in $V^I$ can belong to several clusters, and it is in a cluster $C_k$ with the probability $M_q^{(I|C_k)}$. In addition, a cluster $C_k$ can contain all kinds of entities. So, the approach is a soft clustering.

## IV. METHODS

In this section, we study a probabilistic model to estimate the probability of attributive entities and the target entity. Moreover, the probability of entities can effectively deduce the clustering outcome and represent the significance of entities.

### A. FRAMEWORK OF THE THCluster ALGORITHM

For the TCM-HIN, we select a path-based random walk algorithm to estimate the probability of attributive entities in each cluster $C_k$. Random walk is a well-known and useful technique for measuring the node closeness, it has been widely used in homogeneous networks [23], [24]. However, it is rarely utilized in HIN, because heterogeneous entities and links lead to the problem of how to directly use random walk in HIN. In the TCM-HIN, four types of attributive entities, i.e. symptom, disease, formula and function, are connected to the target entity: herb. Inspired by [25], we utilize the random walk method with specified paths. That is, the random walks are directed among attributive entities and need to pass through the target entity. Therefore, the available probability of an attributive entity can be calculated as the sum of the walkers from other attributive entities walking to it through the target entity. Then we use the available probability of attributive entities to generate the probability of target entity. After that, the EM model is used to estimate the posterior probability of entities. According to probability of entities we can subsequently obtain their ranking in each cluster. The above step is repeated until convergence. The core framework of THCluster is shown in Algorithm 1. The next section describes the algorithm in detail.

### B. AVAILABLE PROBABILITY ESTIMATE OF ENITIES

*1) AVAILABLE PROBABILITY FOR ATTRIBUTIVE ENITY:* We start by estimating the probability of attributive entities. The path-based random walk process is formulated with matrix analysis. We utilize $H^{(A^I, A^J|T, G)}$ to represent the probability transition matrix from $I$ type attributive entity $A^I$ to $J$ type attributive entity $A^J$, passing target type $T$ in the TCM-HIN $G$. It can be calculated as follow:

$$H^{(A^I, A^J|T, G)} = H^{(A^I, T|G)} \times H^{(T, A^J|G)} \quad (1)$$

where $H^{(A^I, T|G)}$ is the transition matrix from $A^I$ to $T$ in $G$. For example, in Fig 1, the transition matrix $H^{(S,D|herb, G)}$ means the transition probability from type symptom to disease through herb on $G$. Hereafter, we adjust the transition matrix among attributive entities when considering the clustering information. The clustering information can

**Algorithm 1:** *THCluster*

**Input**: Cluster number **K** and transition matrix **H**
**Output**: Membership probability $M^{(I|C_k)}$ of entities on each cluster $\{C_k\}_{k=1}^{K}$

1. Description: *detecting herbs clusters on TCM-HIN* ;
2. **Begin:**
3. Randomly initialize the membership probability $M^{(I|C_k)}$
4. **while** *non-convergence* **do**
5.     **for** $C_k$ **do**
6.         Estimate the probability of attributive entities: $Pr(V^{(A)}|C_k)$
7.         Generate the probability of target entity: $P(V^{(T)}|C_k)$
8.         Estimate the posterior probability of entities: $P(C_k|V^{(T)}), Pr(C_k|V^{(A)})$
9.     **end**
10. **end**
11. **End**

be represented by the member matrix of target entity. So the conditional transition matrix from $A^I$ to $A^J$ through $T$ in the cluster $C$ can be defined as follows:

$$H^{(A^I,A^J|T,C_k)} = H^{(A^I,T|G)} \times M^{(T|C_k)} \times H^{(T,A^J|G)} \quad (2)$$

where $M^{(T|C_k)}$ is the member of target entity in cluster $C_k$. Here, we need to point out that the $H^{(A^I,A^J|T,G)}$ is fixed for G, and it is a global transition matrix, compared to conditional transition matrix $H^{(A^I,A^J|T,C_k)}$. In this presented method, the global probability of entity id is an important element to smooth the probability of a target entity (Eq.6 shows more details).

The conditional probability of an attributive type $A^I$ on G and $C_k$ are denoted as $Pr(V^{(A^I)}|G) \in [0,1]^{1\times|A^I|}$ and $Pr(V^{(A^I)}|C_k) \in [0,1]^{1\times|A^I|}$. The probability of one type entity is decided by the available probability from others through the target entity. So the conditional probability of attributive type $A^I$ can be defined as follows:

$$Pr(V^{(A^I)}|G) = \sum_{A^J \in A, A^J \neq A^I} Pr(V^{(A^J)}|G) \times H^{(A^I,A^J|T,G)} \quad (3)$$

$$Pr(V^{(A^I)}|C_k) = \sum_{A^J \in A, A^J \neq A^I} Pr(V^{(A^J)}|C_k) \times H^{(A^I,A^J|T,C_k)} \quad (4)$$

The calculation is an iterative process and $Pr(V^{(A^I)}|C_k)$ is initialized as the even value at the first iteration.

*2) AVAILABLE PROBABILITY FOR TARGET ENITY:* Considering that the target entities are affected by adjacent attributive entities, we study a generative model to estimate the probability of target entities. Compared with attributive entities, the probability of target entity $T$ in G can be calculated as follows:

$$Pr(V_t^T|G) = \prod_{A^I \in A} \prod_{q \in N(t)} Pr(V_q^{(A^I)}|G) \quad (5)$$

where $N(t)$ is the set of neighbors of a target type entity $t$ in $G$. It means the target entity $t$ is decided by the different types of its adjacent attributive entities. After that, we consider the probability of the target entity $t$ in a cluster $C_k$. Similarly, the probability is also generated from its adjacent attributive entities in a cluster $C_k$. Moreover, we use the global probability of target entity $V^T$ to smooth the probability. This is calculated as follows:

$$Pr(V_t^T|C_k) = (1-\theta) * \prod_{A^I \in A} \prod_{q \in N(t)} Pr(V_q^{(A^I)}|C_k) + \theta * Pr(V_t^T|G) \quad (6)$$

where the smoothing parameter $0 \leq \theta \leq 1$ represents the weight of global probabilities. A major goal of $\theta$ is to prevent target entities from accumulating into minority clusters and to improve the clustering accuracy. In addition, the $\theta$ makes the probability change of target entities more steady, which enhance the stability of THCluster. The experiments in the following sections will further validate the importance of the smoothing parameter.

### C. POSTERIOR PROBABILITY FOR ENITY

In order to obtain the members of cluster, we need to estimate posterior probability of entities. The primary purpose of the method is to discover the cluster of herbs. Here, we first calculate the posterior probability of target entities and then the posterior probability of attributive entities is determined via the target entities. Now we look for a way to calculate the posterior probability of target entities $P(C_k|V^T)$. According to Bayesian rules, $P(C_k|V^T) \propto P(V^T|C_k) \times P(C_k)$. Unfortunately, the cluster size $P(C_k)$ is unknown, therefore, we need to estimate an appropriate $P(C_k)$ to balance the cluster size. We choose the $P(C_k)$ by maximizing the likelihood of generating target entities in different clusters. The likelihood of target entities is defined as follows:

$$\log L = \sum_{t \in V^T} \log[\sum_{k=1}^{K} P(V_t^T|C_k) \times P(C_k)] \quad (7)$$

The EM model can be utilized to find the latent $P(C_k)$ by maximizing the log $L$. Initially, we set the $P(C_k)$ with even values and then repeat the **E** step and **M** step to iteratively update the latent cluster probability until the $P(C_k)$ obtains convergence. The E step and M step are respectively shown as Eq.8 and Eq.9.

$$P^n(C_k|V^T) \propto P(V^T|C_k) \times P(C_k) \quad (8)$$

$$P^{n+1}(C_k) = \sum_{t \in V^{(T)}} P^n(C_k|V_t^T) \times \frac{1}{|V(T)|} \quad (9)$$

The next step is to estimate the posterior of attributive entities. The basic idea is that the posterior probability of

a attributive entity comes from its neighbor target entity. We calculate it as follows:

$$P(C_k|V_q^{(A')}) = \sum_{t \in N(q)} P(C_k|V_t^T) \times \frac{1}{|N(q)|} \quad (10)$$

Where $P(C_k|V_q^{(A')})$ is the probability of attributive entity $V_q^{(A')}$ belonging to cluster $C_k$ and $N(q)$ is the neighbor set of attributive entity $V_q^{(A')}$. It means that the posterior probability of attributive entity $V_q^{(A')}$ is the average value of its target neighborhoods.

### D. Time Complexity Analysis

Time complexity of THCluster is composed of the following parts. In the network, the complexity of estimating the distribution of attributive enities is $O(n_1*K*|E^*|*|A|)$, where $|E^*|$ is the number of edges in HIN, $|A|$ is the number of attributive nodes, and $n_1$ is the iteration number and $K$ is the cluster number. The complexity of estimating the distribution of target enities is $O(K*|E_t|)$, where $|E_t|$ is the number of edges of target enities. Then the time complexity of calculating posterior probability for target enities is $O(n_2*K*|P|)$ where $n_2$ is the iteration times, $|P|$ is the number of target enities. Similarly, the posterior probability for attributive enities has the complexity $O(K*|E^*|)$. So the complexity for our THCluster algorithm in the HIN is $O(n_3*K*(n_1*|E^*|*|A|+E_t+n_2*|P|+|E^*|))$ where $n_3$ is the iteration number for clustering adjustment in this HIN.

## V. EXPERIMENTS

In this section, we experimentally evaluate the proposed methods. In order to verify the effectiveness of THCluster, we compare the results with several state-of-art methods on real datasets. We implement our method using iGraph, numpy and scipy packages (https://pypi.python.org/pypi).

### A. DATASETS

In this paper, we use the real datasets ChP, the Chinese Pharmacopoeia 2015 Edition (http://wp.chp.org.cn/en/index.html) and 3K+ TCM clinical cases. The ChP Volume I contains a total of 2598 types of medicinal materials, but does not give them any classifications. We use herb information in Volumes I to set up our experiments. The ChP is a unstructured corpus. The corpus contain various information and we only use herb, symptom, disease, formula and function to build the TCM-HIN. We use NLP tools NLPIR (http://ictclas.nlpir.org/) for data processing. The new normalized dataset includes 613 herbs, 467 symptoms, 768 diseases, 1452 formulas and 359 functions.

### B. CLUSTERING EFFECTIVENESS EVALUATION

#### 1) EVALUATION METRICS:
We score the quality of the clusters using the FVIC (Fraction of Vertices Identified

TABLE I
Clustering accuracy for ChP dataset

|  | k-Means | Spectral | PaReCat | THCluster |
|---|---|---|---|---|
| Herbs | 0.563 | 0.782 | 0.857 | **0.875** |
| symptoms | 0.481 | 0.76 | 0.835 | **0.853** |
| formulas | 0.466 | 0.739 | 0.762 | **0.781** |
| functions | 0.401 | 0.689 | 0.694 | **0.735** |
| diseases | 0.261 | 0.599 | 0.603 | **0.630** |

Correctly) method [25]. It has been widely used in many research projects [25], [26] and is defined as follows:

$$olSet(c, c^*) = \{v|v \in c \wedge v \in c^*\}$$

$$maxolSet(c, C_K) = \max_{c^* \in C_K} \{|olSet(c, c^*)|\} \quad (11)$$

$$FVIC = \sum_{c \in C_F} \frac{maxolSet(c, C_K)}{N}$$

where $C_F$ and $C_K$ represent the found and known clusters, respectively. $c$ and $c^*$ are a cluster in $C_F$ and $C_K$, respectively. $N$ is the number of nodes in the network. FVIC evaluates the average matching degree by comparing each predicted cluster with the most matching real cluster. A higher score indicates a better clustering with respect to the ground truth.

#### 2) QUANTITATIVE EVALUATION:
Several approaches are employed in our experiments for comparison. These include: k-means, a common clustering technique; PaReCat, which has been used to cluster Chinese medical records for the task of patient record subcategorization [5], and Spectral Clustering, which has been used to cluster western medical records for the task of predicting healthcare costs for individuals [27]. In this experiment, we fix the smoothing parameter $\theta$ as 0.2 to study convergence and stability. The FVIC for each approach is shown in Table I.

We can observe that THCluster achieves the best accuracy on all entities. The K-means shows poor performance for diseases and an average performance for others. Because it complements text clustering with semantic information, poor semantic information leads to worse results. Spectral has a respectable result. However, it does not consider the structure of the graph. The results of PaReCat is closest to our results, but it does not use these nodes globally. We have shown that THCluster can indeed improve clustering performance within HINs. In addition, we can gain new knowledge from clusters that mismatch the ground truth. We can recommend prescription as references for physicians or patients. We use clinical cases (http://www.cnki.net/) to prove the potential of this direction. This imitates models that learn from the past clinical cases to provide support for the treatment of future patients.

#### 3) QUALITATIVE EVALUATION:
In order to testify THCluster can efectively cluster herbs into informative categories, we apply results to the experts Chen and Zhang who specialize in Chinese materia medica and clinical TCM, to evaluate the clusterting results. THCluster can identify relationships of different enities, include never directly co-occur in any transaction. We show the top ten herbs and symptoms identified by our method,which is shown in Table II. We can see that our method accurately identifies many meaningful relevant herb and symptom. All of these can be classified as stomach meridian in TCM, which were testified by TCM experts, and many of them are widely used in clinical diagnosis.

TABLE II
Top ten herbs and symptoms in a cluster

| Herbs | probability | Symptoms | probability | Rank |
|---|---|---|---|---|
| pinellia ternata(半夏) | 0.679 | epigastric abdominal pain(脘腹胀痛) | 0.862 | 1 |
| fructus aurantii(枳壳) | 0.653 | chest abdominalL distension(胃脘痞满) | 0.86 | 2 |
| fructus hordei germinatus(麦芽) | 0.652 | eructation(嗳气) | 0.857 | 3 |
| et corneum gigeriae galli(鸡内金) | 0.642 | not hungry(不饥食少) | 0.763 | 4 |
| pericarpium citrus reticulata(陈皮) | 0.635 | sallowcomplexion(面色萎黄) | 0.351 | 5 |
| vladimiria souliei(川木香) | 0.532 | dry stool(大便不爽) | 0.345 | 6 |
| codonopsis pilosula(党参) | 0.513 | dry mouth(口干) | 0.256 | 7 |
| fructus toosendan(川楝子) | 0.432 | emesis(呕吐) | 0.231 | 8 |
| paeoniae alba(白芍) | 0.412 | Fatigue(乏力) | 0.217 | 9 |
| glycyrrhiza uralensis(甘草) | 0.367 | chest tightness(胸闷) | 0.203 | 10 |

TABLE III
Comparison between THCluster and TCNM experts in Prescription

| Symptoms | Approaches | Prescription | Common herbs |
|---|---|---|---|
| epigastric pain(胃脘痛), vomiting(呕吐), eructation(嗳气), dry stool(大便干), redeye(目赤), headache(头痛) | THCluster | pinellia ternata(半夏), codonopsis pilosula(党参), vladimiria souliei(川木香), et corneum gigeriae galli(鸡内金), fructus aurantii(枳壳), pericarpium citrus reticulata(陈皮) | pinellia ternata(半夏), codonopsis pilosula(党参), vladimiria souliei(川木香), et corneum gigeriae galli(鸡内金), fructus aurantii(枳壳), pericarpium citrus reticulata(陈皮), fructus hordei germinatus(麦芽) |
| | Expert1 | pinellia ternata(半夏), codonopsis pilosula(党参), vladimiria souliei(川木香), et corneum gigeriae galli(鸡内金), fructus aurantii(枳壳), radix paeoniae alba(白芍), glycyrrhiza uralensis(甘草) | |
| | Expert2 | pinellia ternata(半夏), vladimiria souliei(川木香), et corneum gigeriae galli(鸡内金), fructus aurantii(枳壳), radix paeoniae alba(白芍), pericarpium citrus reticulata(陈皮) | |

*4) CASE EVALUATION:* We use three standard metrics, $F_1$ score, Precision and Recall, to evaluate the herb prediction task. We take the macro-average of these three metrics (Figure 2). We compare our method with PaReCat, which is a novel and effective approaches in TCM clinical cases mining (Figure 2). We see that our method is slightly better than the PaReCat method on all metrics for herb prediction. We note that prescription prediction based on symptoms and diseases, achieves a reasonably good scores for $F_1$, precision, and recall. This suggests that the system can identify the herbs that the physician actually prescribed to a patient, whilst also recommending many herbs that were not prescribed.

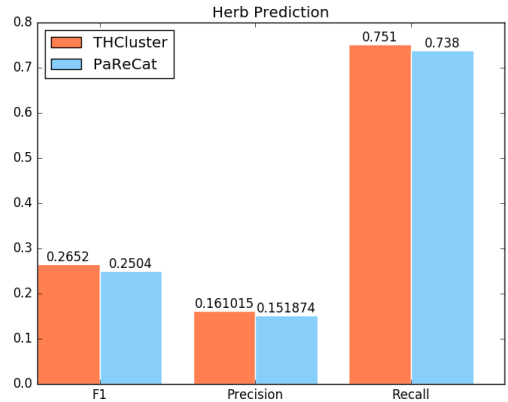

Fig. 2. A comparison between THCluster and PaReCat for herb classification.

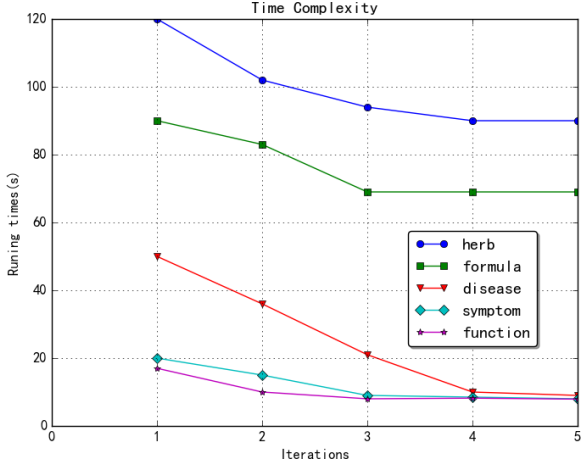

Fig. 3. The time of analyzing TCM-HIN along with the iterations.

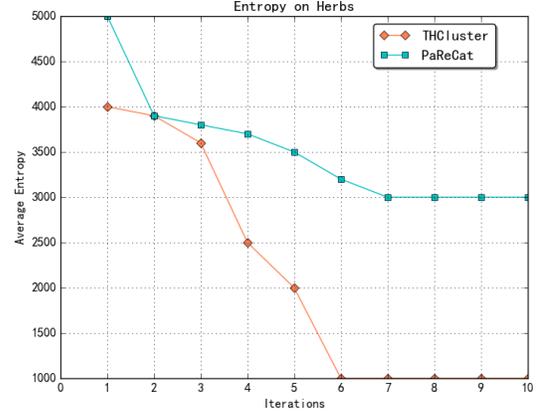

Fig. 4. The change of averge entropy with iterations.

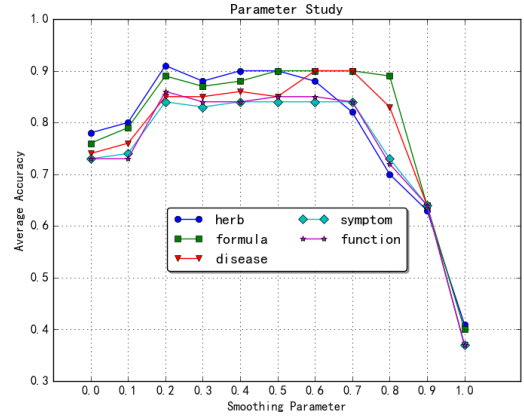

Fig. 5. The change of averge entropy with smoothing parameter $\theta$.

To further verify our results in detail, we compare herbs recommended by our method to herbs prescribed by TCM experts for a patient suffering from chronic gastritis and enteritis (Table III). We can see that our model prescribed seven herbs in common with the thirteen prescribed by the physicians. Our model also recommends one herb not prescribed by the physicians. A physician verified that these seven herbs are all known to be associated with curing either constipation or chronic gastritis.

*5) Time Complexity Study:* We analyze the running time of TCM-HIN along with the iterations on ChP dataset. The results are shown in Fig 3. The complex network cost much more running time. It is reasonable, since more nodes and links need to be handled in this condition. In addition, the analysis time of network decreases along the iteration. Because the prior knowledge inherited from previous iterations on the network helps to fasten convergence. Although the iteration process in TCM-HIN results in its higher time complexity, the time used in each iteration drops down quickly.

*6) Convergence and Stability Study:* We study the convergence and stability of THCluster on the ChP dataset. The entropy is used to evaluate the stability of a cluster and the convergence of a algorithm. A better method has a lower entropy. The average entropy is defined as follows:

$$Entropy(V^{(I)}) = -\frac{1}{K}\sum_{k=1}^{K^{(I)}}\sum_{t=1}^{|V^{(I)}|} P(C_k|V_t^{(I)}) \log P(C_k|V_t^{(I)}) \quad (12)$$

The comparison of average entropy of THCluster and PaReCat on herbs of TCM-HIN is shown in Fig 4. It is obvious that THCluster achieves lower averge entropy. The reason is that the THCluster rationally combines more information from all types of enities. It helps THCluster to achieves steady performance.

*7) Parameter Study:* There is a smoothing parameter $\theta$. We use clustering accuracy for different types of enities to analyze the effect of different parameter $\theta$. The smoothing parameter $\theta$ is used to control the portion of global probability used by each cluster. We run THCluster on Chp dataset with different $\theta$. The result are shown in Fig 5. We can discover that THCluster achieves better results when $\theta$ is from 0.2 to 0.5. It imples that the appropriate global information is helpful for clustering. Too much( =1) or no global information($\theta$ is small) both will drop down the performance of THCluster.

## VI. Conclusion

Mining categorizations of herbs for the precision prescription in TCM is an essential and beneficial for precision medicine. In this paper, we studied a novel clustering model called THCluster. Based on HIN, a path-based random walk model is utilized in THCluster to estimate the available probability of an entity. Meanwhile, THCluser uses

Bayesian rules and EM models to estimate the posterior probability of entities. THCluster is a new probabilistic model that can be effectively used for clustering analysis. The experiments show that THCluster achieves better clustering results than other representative algorithms. We performed experiments on real-world datasets and observed an improvement in herb categorization, which have been authenticated by TCM experts. THCluster is a general solution and can be used to solve other problems. In our future work, we will discover more latent knowledge in various bio-medicine datasets.

To further improve our method, we will study deep-walk to estimate probability of entities. Moreover, the current model is developed for relationships of herb analysis. Another important future work is to perform Link Prediction on the TCM-HIN to discover latent relationships among different types of entities.

## Acknowledgment

This work is supported in part by the Chinese National Natural Science Foundation(No.61332013), the Research Innovation Plan of Shanghai Education Commission(No.14YS026) and Research Foundation of Shanghai Municipal Health and Family Planning Commission(No.JP013).